\documentclass[twocolumn]{aastex62}
\usepackage{amsbsy}
\usepackage{multirow}

\def\kms{{\,\rm km\,s^{-1}}}
\def\msun{\,{\rm h^{-1} M}_\odot}

\def\mpch{\,{\rm h^{-1}  Mpc}}
\def\kpch{\,{\rm h^{-1}  kpc}}
\def\lsim{\mathrel{\lower0.6ex\hbox{$\buildrel {\textstyle <}
 \over {\scriptstyle \sim}$}}}
\def\gsim{\mathrel{\lower0.6ex\hbox{$\buildrel {\textstyle >}
 \over {\scriptstyle \sim}$}}}
\def\lone{\lambda_{1}}
\def\ltwo{\lambda_{2}}
\def\lthree{\lambda_{3}}
\def\bx{\,{\boldsymbol{x}}}

\def\e{\,{\boldsymbol{e_1}}}
\def\ee{\,{\boldsymbol{e_2}}}
\def\eee{\,{\boldsymbol{e_3}}}

\received{August 20, 2018}
\revised{September 7, 2018}
\accepted{September 14 2018}

\shorttitle{galaxy spin--LSS alignment}
\shortauthors{Peng Wang et al.}

\begin{document}

\title{The spin alignment of galaxies with the large-scale tidal field in hydrodynamic simulations}

\correspondingauthor{Peng Wang}
\email{wangpeng@pmo.ac.cn}

\author{Peng Wang}
\affil{Purple Mountain Observatory (PMO), No. 8 Yuan Hua Road, 210034 Nanjing, China}
\affil{Graduate School, University of the Chinese Academy of Science, 19A, Yuquan Road, Beijing 100049, China}

\author{Quan Guo}
\affil{Shanghai Astronomical Observatory (SHAO), Nandan Road 80, Shanghai 20030, China}

\author{Xi Kang}
\affil{Purple Mountain Observatory (PMO), No. 8 Yuan Hua Road, 210034 Nanjing, China}

\author{Noam I. Libeskind}
\affil{Leibniz-Institut f\"ur Astrophysik Potsdam (AIP), An der Sternwarte 16, 14482 Potsdam, Germany}


\begin{abstract}
The correlation between the spins of dark matter halos and the large-scale structure (LSS) has been studied in great detail over a large redshift range, while investigations of galaxies are still incomplete. Motivated by this point, we use the state-of-the-art hydrodynamic simulation, Illustris-1, to investigate mainly the spin--LSS correlation of galaxies at redshift of $z=0$. We mainly find that the spins of low-mass, blue, oblate galaxies are preferentially aligned with the slowest collapsing direction ($\eee$) of the large-scale tidal field, while massive, red, prolate galaxy spins tend to be perpendicular to $\eee$. The transition from a parallel to a perpendicular trend occurs at $\sim10^{9.4}\msun$ in the stellar mass, $\sim0.62$ in the g–r color, and $\sim0.4$ in triaxiality. The transition stellar mass decreases with increasing redshifts. The alignment was found to be primarily correlated with the galaxy stellar mass. Our results are consistent with previous studies both in N-body simulations and observations. Our study also fills the vacancy in the study of the galaxy spin--LSS correlation at $z=0$ using hydrodynamical simulations and also provides important insight to understand the formation and evolution of galaxy angular momentum.
\end{abstract}

\keywords{
methods: statistical ---
methods: observational ---
galaxies: evolution ---
galaxies: general ---
cosmology: large-scale structure of Universe.
}

\section{Introduction}\label{sec:intro}

How the spin of a halo/galaxy is correlated with the large- scale cosmic-web environment is an important question to address for understanding both galaxy formation and the intrinsic alignment of galaxies in the context of weak gravitational lensing. The tidal torque theory (TTT) suggested that there is an alignment between the halo/galaxy spin and the large-scale structure \citep[LSS,][]{1951pca..conf..195H, 1969ApJ...155..393P, 1987ApJ...319..575B, 1984ApJ...286...38W}, which is often referred to as the spin--LSS correlation. However, the TTT predictions on small scales are affected by the nonlinear evolutions, which may significantly influence the way of mass infall onto halo/galaxy.

In the very first study of N-body simulation by \cite{2007ApJ...655L...5A}, who applied a cosmic-web classification called a multiscale morphology filter \citep[MMF,][]{2007A&A...474..315A}, they showed that the spin--LSS correlation depends on the halo mass, i.e., the spin of low-mass halos with masses less than $10^{12}\msun$ are found to be parallel with the direction of the filamentary structure, while a perpendicular signal was found for high-mass halos. Shortly after, \cite{2007MNRAS.381...41H} examined halos in two mass bins (low mass: $5\times10^{10}-10^{12}\msun$; high-mass: $>10^{12}\msun$) and claimed a weak anti-alignment for halos in filaments and a stronger anti-alignment in sheets. With the development of numerical simulations and the improvement of cosmic-web classification \citep[e.g., P-web, T-web and V-web, a review see][]{2018MNRAS.473.1195L}, many studies using dark matter simulations have confirmed the spin--LSS correlations \citep{2009ApJ...706..747Z, 2013MNRAS.428.2489L, 2014MNRAS.443.1274L, 2013ApJ...762...72T, 2014MNRAS.443.1090F, 2014MNRAS.440L..46A, 2016IAUS..308..421P, 2017MNRAS.468L.123W, 2018MNRAS.473.1562W, 2018arXiv180500033G}.

The above studies have found that the transition halo mass of the spin--LSS correlation ranges from $5\times10^{11}\msun$ to $5\times10^{12}\msun$, which weakly depends on the mass resolution of the simulation and the smoothing scales used to classify the cosmic web. \cite{2012MNRAS.427.3320C} and \cite{2016IAUS..308..421P} proposed an empirical formula, where $\rm M_{trans}^{z}=M_{trans}^{0}(1+z)^{-\gamma_{s}}$ with $\gamma_{s}=2.5\pm0.2$ and where $\rm M_{trans}^{0}$ is the transition mass equal to $\rm 5(\pm1)\times10^{12}M_{\odot}$.

In addition to those works based on N-body simulations, using the hydrodynamic simulation Horizon-AGN, \cite{2014MNRAS.444.1453D} investigated the spin– LSS correlation as functions of galaxy properties (such as the stellar mass, color, and star formation rate) in the redshift range of $1.2<z<1.8$. They claimed that low-mass blue galaxies are preferentially aligned with their nearest filaments and that high-mass red galaxies show a tendency for a perpendicular signal. They found that the transition mass from alignment to misalignment happens at $\sim3\times10^{10}\msun$.

These theoretical predictions have to be justified by observations. Using the observational data from Sloan Digital Sky Survey (SDSS) DR8 and the filament finder of the Bisous process \citep{2014MNRAS.438.3465T}, \cite{2013ApJ...775L..42T,2013MNRAS.428.1827T} confirmed that such a correlation depends on the galaxy type, i.e., the spin of spiral galaxies tends to align with their nearest filaments, while the short axes of ellipticals are perpendicular to the filament. \cite{2015ApJ...798...17Z} claimed a mass-dependent correlation where the spins of spiral galaxies have the weak tendencies to be aligned with (or perpendicular to) the intermediate (or minor) axis of the local tidal tensor. \cite{2016MNRAS.457..695P} used the galaxy sample constructed from the 2 Micron All-Sky Survey (2MASS) Redshift Survey \citep{ 2012ApJS..199...26H} and examined the alignment between the galaxy spin and the velocity shear field \citep[V-web, ][]{2012MNRAS.425.2049H, 2013MNRAS.428.2489L}. They reported a significant perpendicular signal with respect to the axis of the slowest compression for elliptical galaxies, while no signal was found in spiral galaxies.

A summary of previous investigations is shown in Table.~\ref{table:table1}. Clearly, studies using hydrodynamical simulations and observational data are lagging behind works that use N-body simulations.

As can be seen, most works have used N-body simulations, and a few have used hydrodynamic simulations and observations. The measurement of a galaxy's spin and the cosmic-web classification is difficult to observe, and only a few studies are accomplished at a low redshift. Limited by the resolution and cost of hydrodynamics simulations, only a few works investigated the spin--LSS correlation of a galaxy at a high redshift. Thanks to the advent of cosmological hydrodynamics simulations, such as Illustris project \citep{2014MNRAS.444.1518V}, we are able to extend the galaxy spin--LSS correlation to the present time. This is the main motivation of this work.

The structure of this paper is organized as follow. In Section~\ref{sec:method} we will introduce the simulation data, the spin definition, and the cosmic-web classification. In Section~\ref{sec:results} we will show the main results. A discussion and the conclusion are presented in Section~\ref{sec:con_dis}.

\begin{deluxetable*}{c|ccc}
\tablecaption{Summary of previous works on the spin--LSS correlation.} 
\label{table:table1}
\tablehead{
\colhead{} &
\colhead{Author} & 
\colhead{Sample} & 
\colhead{Redshift}
}
\startdata
\multirow{4}{*}{\textbf{Hydrodynamic}}
& Wang et al. 2018 & Galaxy & z=0, 1, 2 \\
& (This Work) & & \\
& \cite{2014MNRAS.444.1453D} & Galaxy & $1.2<z<1.8$  \\
& \cite{2012MNRAS.427.3320C} & Galaxy & $1.2<z<1.8$  \\
\hline
\multirow{8}{*}{\textbf{N-body}}
& \cite{2018arXiv180500033G} & Halo & z=0  \\
& \cite{2017MNRAS.468L.123W, 2018MNRAS.473.1562W}  & Halo & $z\leq5$   \\
& \cite{2014MNRAS.443.1090F} & Halo &  $z=0$ \\
& \cite{2013ApJ...762...72T} & Halo & $z=0,1,2,3$  \\
& \cite{2013MNRAS.428.2489L} & Halo & $z=0$ \\
& \cite{2009ApJ...706..747Z} & Halo & $z=0$  \\
& \cite{2007MNRAS.381...41H} & Halo & $z=0, 0.49, 1.05$ \\
& \cite{2007ApJ...655L...5A} & Halo & $z=0$ \\
\hline
\multirow{3}{*}{\textbf{Observations}}
& \cite{2016MNRAS.457..695P} & Galaxy & low redshift \\
& \cite{2015ApJ...798...17Z} & Galaxy & low redshift \\
& \cite{2013ApJ...775L..42T} & Galaxy & low redshift \\
& \cite{2013MNRAS.428.1827T} & Galaxy & low redshift \\
\enddata
\end{deluxetable*}

\section{Data and Methodology}\label{sec:method}
The data used in this work were constructed from the state- of-the-art hydrodynamic simulations, Illustris-1\footnote{\url{http://www.illustris-project.org/data/}}, which is the one with highest resolution simulations and consists of full physics in the Illustris project \citep{2014MNRAS.444.1518V}. The Illustris project consists of a set of cosmological hydrodynamic simulations, which are applied by the moving mech code, AREPO \citep{2010MNRAS.401..791S}. The cosmological parameters are consistent with the Wilkinson Microwave Anisotropy Probe (WAMP)-9 measurements \citep{ 2013ApJS..208...19H}, namely: $\Omega_m=0.2726$, $\Omega_\Lambda=0.7274$, $\Omega_b=0.0456$, $\sigma_8=0.809$, $n_s=0.963$ and $H_0=100 \ h \kms$ with $h=0.704$. The Illustirs-1 simulation evolves $1820^{3}$ dark matter particles and the same number of gas cells from $z=127$ to $z=0$ in a volume of a $\rm 75 \mpch$ wide periodic cosmological box. The mass resolution is $6.26\times10^6 \msun$ in dark matter and is $1.26\times10^6 \msun$ in baryonic matter.

The standard Friend-of-Friend (FoF) algorithm \citep{1985ApJ...292..371D} is used to identify dark matter halos. The linking length is set to 0.2 times of the mean particles separation. In total,  7713,601 FOF groups with more than 32 particles are found in Illustris-1 at $z=0$. Stellar particles and gas cells were attached to these FOF groups in a secondary linking stage \citep{2009MNRAS.399..497D}. The SUBFIND algorithm \citep{2001MNRAS.328..726S, 1985ApJ...292..371D} is applied for every FoF group to identify gravitationally bound structures. The halo/galaxy positions are defined as the position of the most bound particle of the biggest structure. Galactic properties (such as stellar mass, light, color, etc.) are measured within the radius, $r_{\star}$, which is equal twice the stellar half-mass radius of each SUBFIND (sub)halo.

In order to make a robust measurement of the galaxy spin, we selected 36,327 galaxies with at least $300$ star particles within a radius of $30\kpch$, (for comparison, another scale of $15\kpch$ is also considered in our work). The galaxy spin is defined as
\begin{equation}
\boldsymbol{j}=\sum_{i=1}^{N}m_{i}\boldsymbol{r_{i}}\times(\boldsymbol{v_{i}}-\boldsymbol{\bar{v}})
\end{equation}
in which $N$ is the total number of star particles within $30\kpch$ ($15\kpch$), $m_{i}$ is the mass of the star particle $i$ and $\boldsymbol{r_{i}}$ is the position vector of the star particle $i$ relative to the galaxy center.  $\boldsymbol{v_{i}}$ is the velocity of the star particle $i$ and $\boldsymbol{\bar{v}}$ is the mean velocity of all star particles within the given scale.

To determine the direction of the LSS around each galaxy, all of the particles (including dark matter, star, and gas particles) are considered. We calculate the Hessian matrix of the smoothed tidal tensor field at the galaxy center, which is define as
\begin{equation}
T_{ij}(\bx) = \frac {\partial^2 \phi} {\partial x_i \partial x_j},
\label{TidalTensor}
\end{equation}
in which $i, j = 1, 2, 3$ are indices representing spatial dimensions. $\phi$ is the peculiar gravitational potential. By following the commonly used value suggested in previous works on the studies of the halo spin--LSS relation \citep[e.g.][]{2007ApJ...655L...5A, 2007MNRAS.381...41H, 2009ApJ...706..747Z, 2012MNRAS.427.3320C, 2013ApJ...762...72T}, we set the smoothing length as $\rm R_{s} = 2 \mpch$. Notes that in our previous works \citep{2017MNRAS.468L.123W, 2018MNRAS.473.1562W}, we compared several values of $\rm R_{s}$ and it was found that there is no significant effect on halo spin--LSS correlation. The tidal tensor is subjected to a principle component analysis, and its eigenvalues ($\lone \ge \ltwo \ge \lthree$) with corresponding eigenvectors ($\e,\ee,\eee$) are computed. According to the seminal work of the \cite{1970A&A.....5...84Z}, the eigenvector $\eee$ (corresponding to $\lthree$) defines the slowest compression of mass on a large scale. For instance, for a filamentary environment, $\eee$ is the filament direction, and for a sheet like environment,  $\eee$ lies in the sheet plane. In the following analysis, we refer to $\eee$ as the direction of the large-scale tidal field.

To quantify the correlation between the spin of a galaxy with the direction of the LSS, we compute the $\cos(\theta)=\boldsymbol{j} \cdot \eee$ between the galaxy spin vector ($\boldsymbol{j}$) and $\eee$.  Note that we restrict $\cos(\theta)$ to be within $[0,1]$. In the case where the galaxy spin is randomly oriented relative to $\eee$, the expectation of  $\langle|\cos(\theta)|\rangle$ is 0.5; and if $\langle|\cos(\theta)|\rangle$ larger than 0.5, we refer to it as an alignment between the galaxy spin and the $\eee$; if $\langle|\cos(\theta)|\rangle$ smaller than 0.5, it means the galaxy spin is preferentially perpendicular to $\eee$.

 \section{Results}\label{sec:results}
\begin{figure*}
\plotone{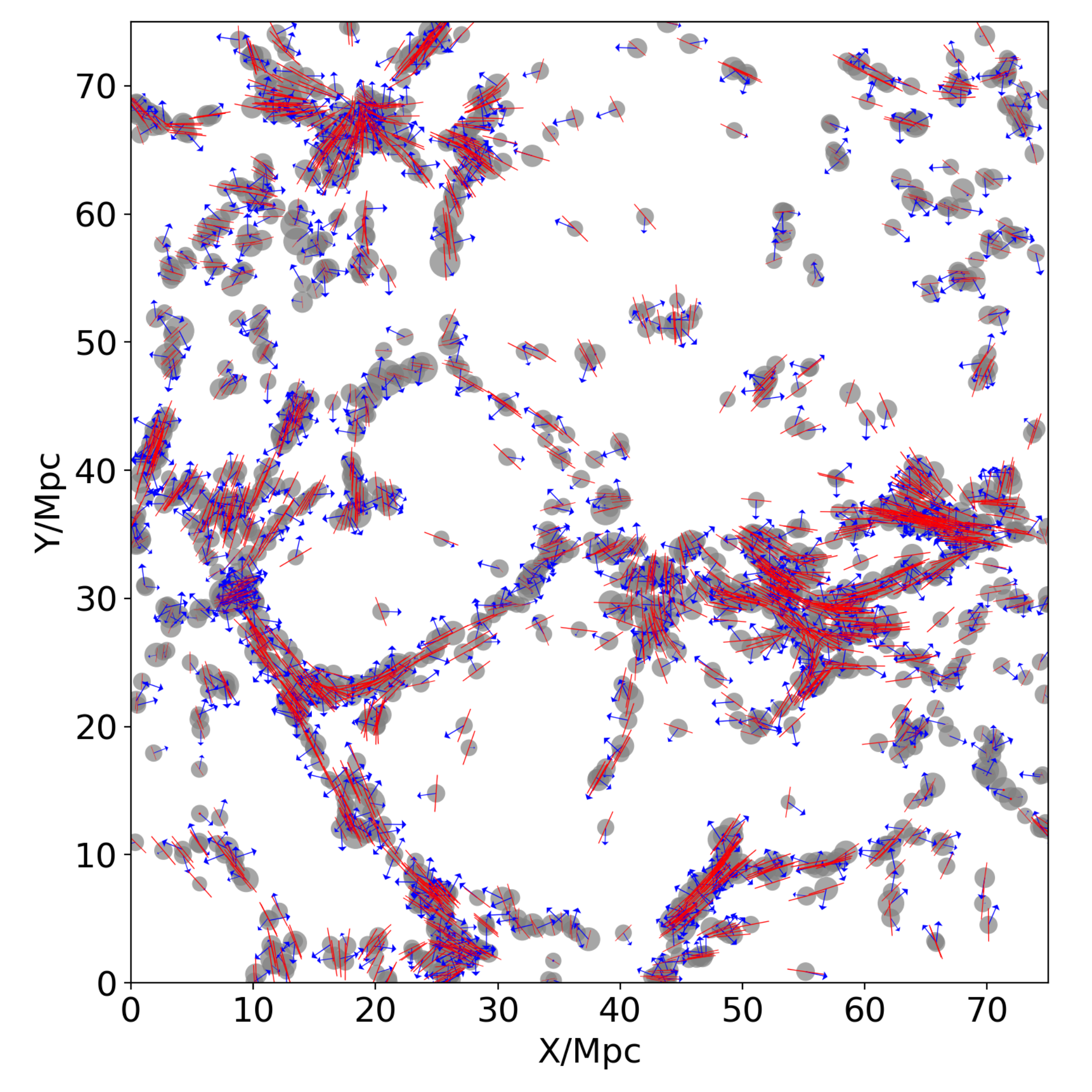}
\caption{Illustration of the correlation between the galaxy spin and the direction of the large-scale tidal field from the Illustris-1 simulation in a slice of width that is $2\mpch$ across the z axis. Gray circles show galaxies with at least 300 star particles within $30\kpch$. The red arrows indicate the direction of the large-scale tidal field, $\eee$. The blue arrows indicate the direction of the galaxy spin.}
\label{fig:f1}
\end{figure*}

In Fig.~\ref{fig:f1}, we show the space configuration of the spin-$\eee$ alignment of the galaxies in Illustris-1 at $z=0$. For clarity, we only show a slice of width that is $2 \mpch$ across the z axis and galaxies with at least 300 star particles within $30\kpch$ (gray circles). Red arrows represent the slowest collapse direction, $\eee$, of the large-scale tidal field. Blue arrows indicate galaxy spins. Generally speaking, the distribution of galaxies shows some filament-like and knot-like structures, and the red arrows generally point along with the filament-like structures and are radially aligned with the knot-like structures. Galaxy spins (blue arrows) show two kinds of trends with respect to the $\eee$: either parallel or perpendicular.

\begin{figure}[!ht]
\plotone{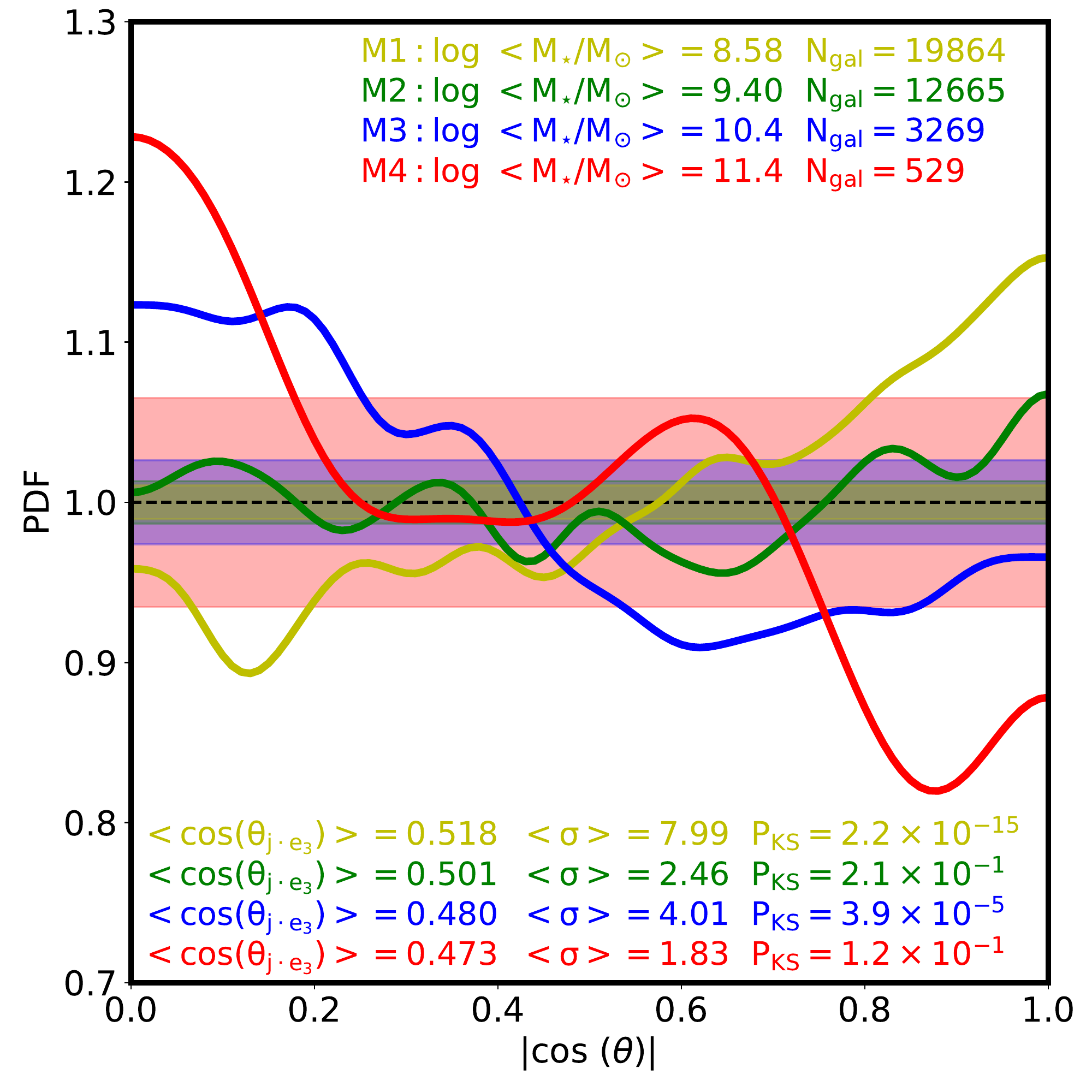}
\caption{Probability distribution of the $\cos(\theta)$, where $\theta$ is the angle between the galaxy spin (calculated within $30 \kpch$) and the direction of the large-scale tidal field, $\eee$. The horizontal black dashed line represents a random distribution. Solid color lines shows the measured alignment signal for galaxies in the four mass bins. The color filled regions show the spread of the $3\sigma$ deviation from the 10,000 random uniform distribution of the same size for each subsample. The mean values of $\cos(\theta)$, the significance, and $p_{KS}$ of the Kolmogoro--Smirnov (KS) test are shown in the bottom-left corner with different colors, respectively. The Kolmogoro--Smirnov (KS) test was performed to quantify the likelihood that these are consistent with being derived from a uniform distribution.}
\label{fig:f2}
\end{figure}

To quantify the spin--LSS correlation, Fig.~\ref{fig:f2} shows the probability distribution of the $\cos(\theta)$, in which $\theta$ is the angle between the galaxy spin measured using star particles within $30\kpch$ and $\eee$ of the tidal field. Galaxies are divided into four bins according to their stellar mass. The upper and lower limit of each bin is M1=[min($M_{\star}$), $10^{9}$]; M2= [$10^{9}$, $10^{10}$]; M3=[$10^{10}$, $10^{11}$]; and M4= [$10^{11}$,max($M_{\star}$)]. The mean value of each bin is: $10^{8.58}$ (M1, yellow), $10^{9.40}$ (M2, green), $10^{10.4}$ (M3, blue) and $10^{11.4}$ (M4, red), respectively, in units of $\msun$. The black dashed line means that the spin is randomly distributed with the $\eee$. The color regions (with respect to the sample with same color) show the $3\sigma$ spread constructed from 10,000 random uniform distributions with the same size for each sample. To construct the random samples, we kept the $\eee$ fixed and randomized the direction of the galaxy spin. The width of the color regions is inversely proportional to the sample size. In the bottom-left corner, we show the mean value of $\cos(\theta)$, the mean significance $\langle\sigma\rangle$, and the p-value of the Kolmogorov--Smirnov (KS) test, $p_{\rm KS}$. The mean significance and the KS test are performed on the full set of angles to quantify the likelihood that these are consistent with being derived from a uniform distribution. Note that a high value of significance and a low value of $p_{\rm KS}$ indicates a significant signal.

\begin{figure}[!ht]
\plotone{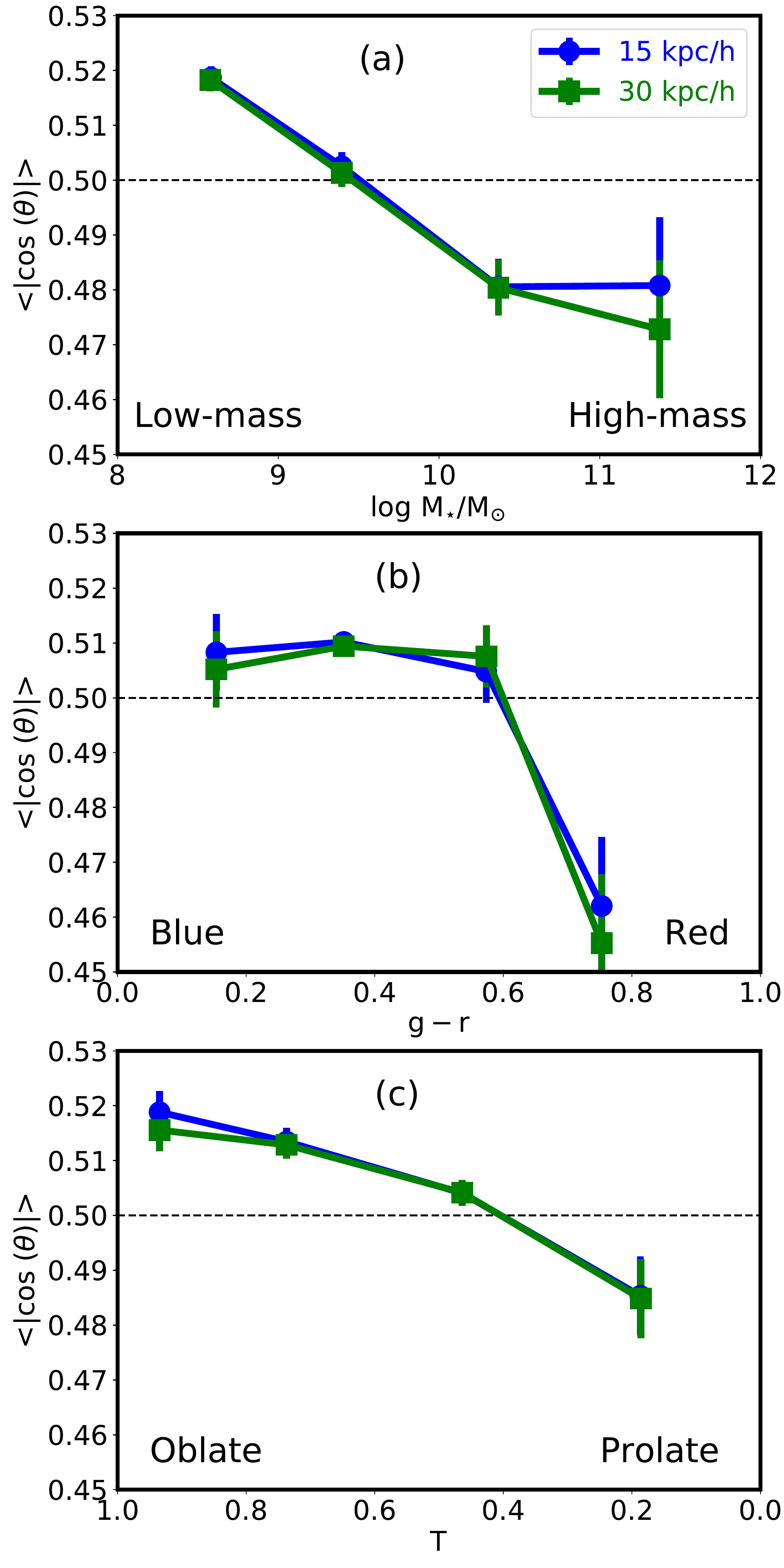}
\caption{Alignment between the galaxy spin and the direction of the large- scale tidal field, e3, as function of the galaxy property: stellar mass (top panel), $\rm g-r$ color (middle panel), and 3D shape triaxiality, $\rm T$ (bottom panel). Blue (green) lines with error bars show the galaxy spin calculated within $30\kpch$ ($15\kpch$). Error bars show Poisson errors. Note that the horizontal dashed line corresponds to a random distribution betweten the galaxy spin and $\eee$ of the large-scale tidal field.}
\label{fig:f3}
\end{figure}

We see that a relatively strong alignment signal is found for low--mass galaxies in M1 (yellow line) with a mean value of $\cos(\theta)=0.518$, a high value of significance ($\sim8.0\sigma$), and $p_{\rm KS}=2.2\times10^{-15}$. The large number of galaxies in this sample results in the high value of significance and the low value of $p_{\rm KS}$. The alignment signal becomes almost randomly distributed in the M2 (green line) sample with a mean value of $\cos(\theta)=0.501$. However, the two massive samples (M3 in blue and M4 in red) show perpendicular trends, with $\cos(\theta)$ equal to 0.480 and 0.473, respectively. We found that the number of galaxies in those two sample is relative small, leading to a relative small significance and a high $p_{\rm KS}$.

That the strength of the alignment (the value of $\cos(\theta)$) decreases with galaxy the stellar mass, which indicates that the spin--LSS correlation of a galaxy changes from a parallel trend at the low-mass end to a perpendicular trend at the high-mass end. This generally agrees with previous simulation works \citep[e.g.,][]{2007ApJ...655L...5A, 2007MNRAS.381...41H, 2013ApJ...762...72T, 2014MNRAS.443.1090F, 2017MNRAS.468L.123W, 2018MNRAS.473.1562W}, It also agrees, to some extent, with some observational results \citep[e.g.,][]{ 2013ApJ...775L..42T} under the assumption that spiral galaxies are mainly hosted by low--mass halos and elliptical galaxies are mainly hosted by relatively massive halos.

\begin{figure}[!ht]
\plotone{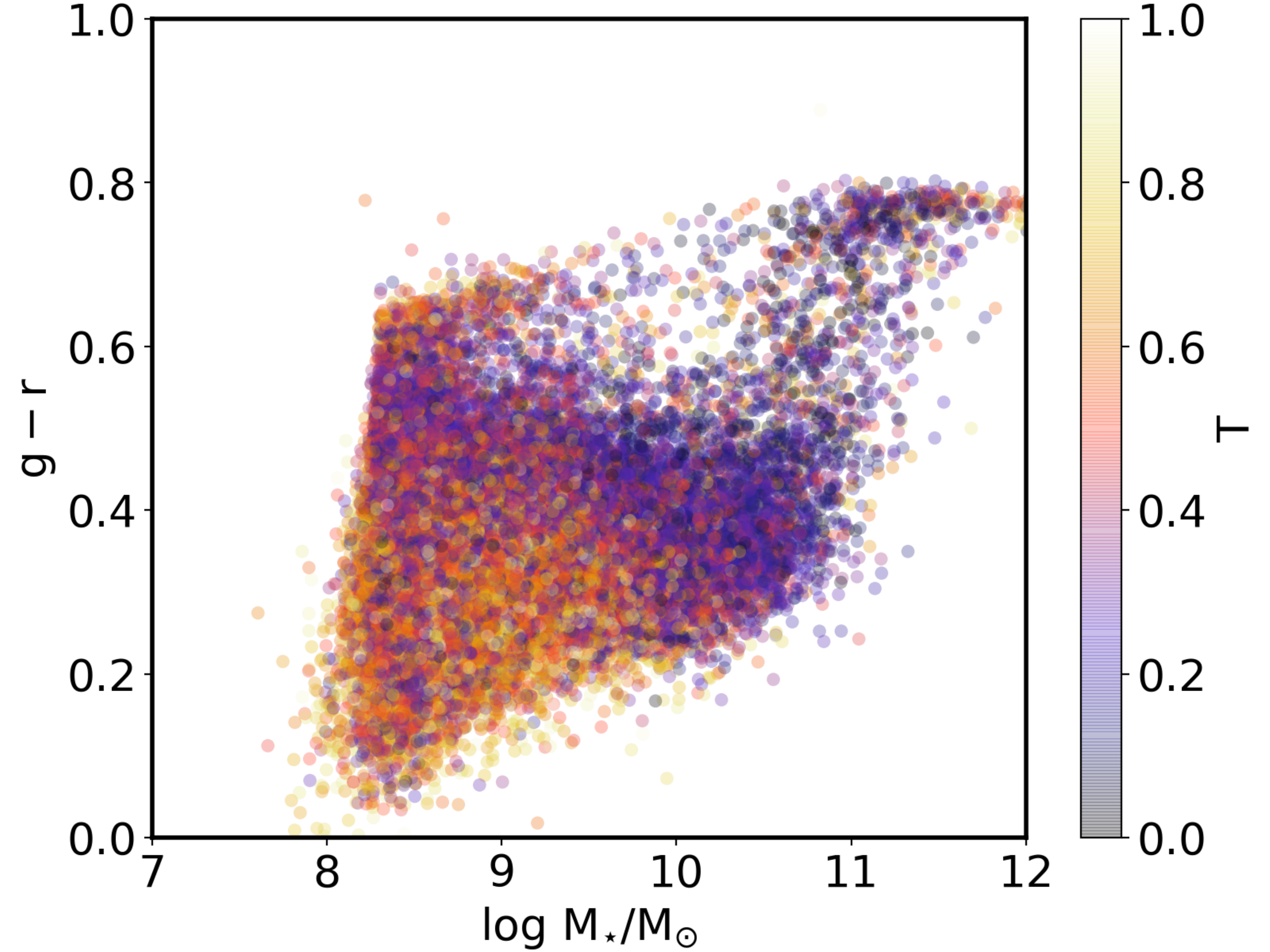}
\caption{Distribution of galaxy properties: stellar mass, $g-r$ color, and the triaxiality.}
\label{fig:f4}
\end{figure}

In order to compare with observational results, we show the impact of galaxy properties (mass, $g-r$ color, and triaxiality) on the galaxy spin--LSS correlation in Fig.~\ref{fig:f3}. As a comparison, two scales ($\rm 15 \kpch$ in blue  and $\rm 30 \kpch$ in green) were used to measure the galaxy spin. Error bars in each panel show the Poisson errors. In panel-(a), we show the $|\cos(\theta)|$ as a function of galaxy stellar mass, which is similar to Fig.~\ref{fig:f2}. Panel-(b) shows the g-r color dependence. Galaxies with low values of g–r are labeled ``blue'', while high values of g–r are labeled as ``red''. In panel-(c), a galaxy's triaxiality is computed by $\rm T=\frac{a^2-b^2}{a^2-c^2}$, in which $a\geq b\geq c$ are the axes of the galaxy correspondingly measured in the same scale. Purely prolate galaxies have $\rm T=1$, while purely oblate galaxies have $\rm T=0$.

\begin{figure*}[!ht]
\plottwo{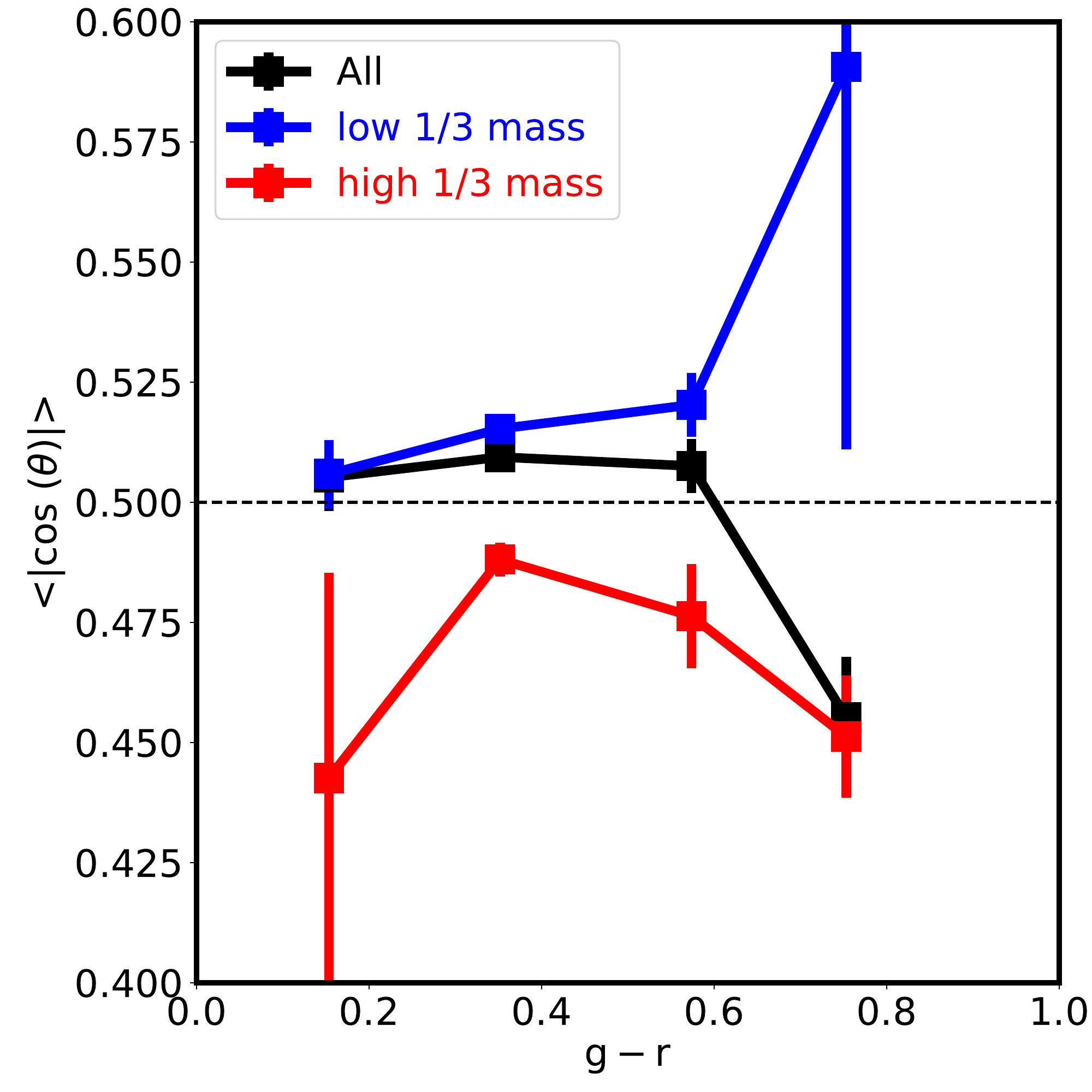}{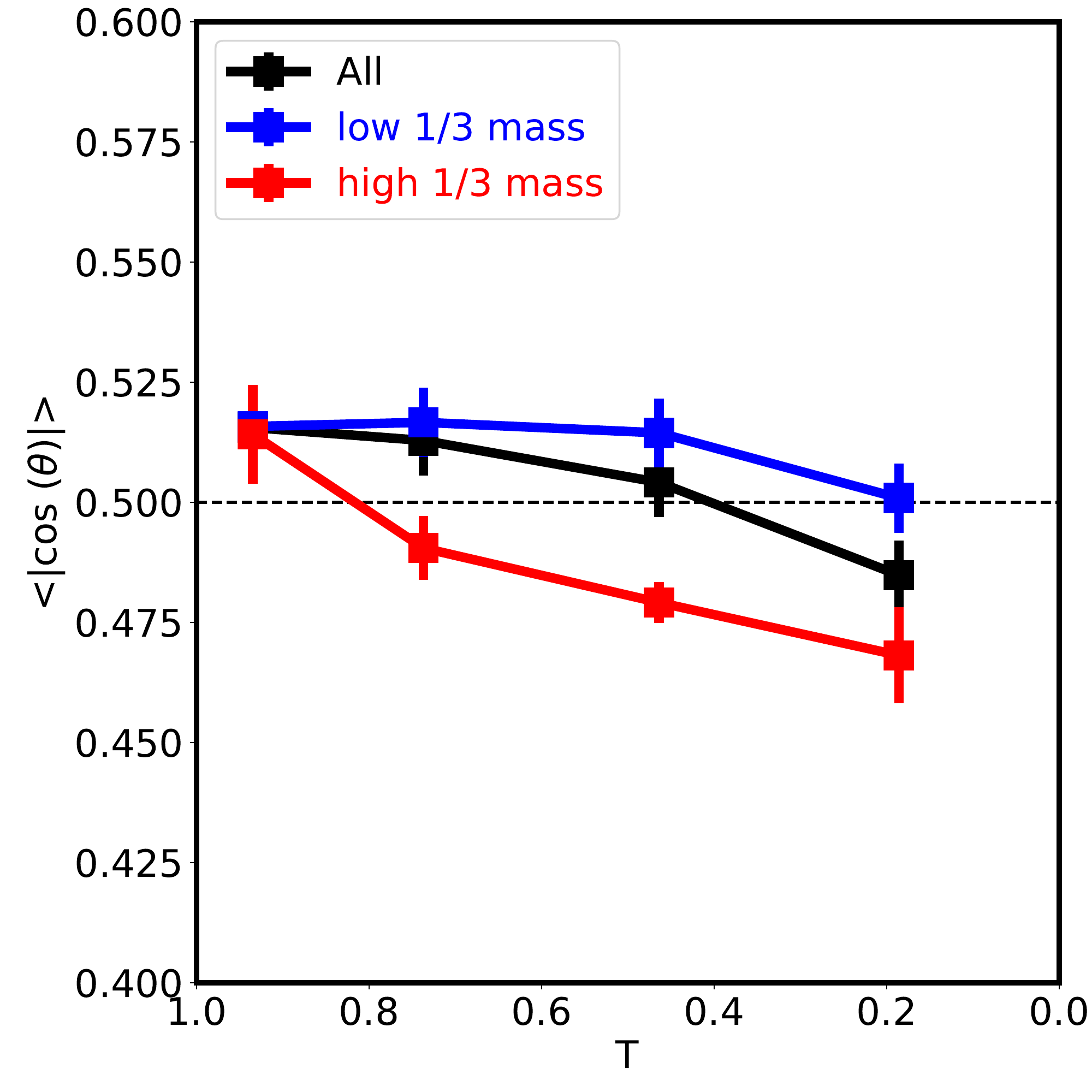}
\caption{Similar to Fig.~\ref{fig:f3}, but galaxies at each color bin (left panel) and shape triaxiality (right panel) are further divided into three bins by their stellar mass. The blue (red) lines represent for galaxies with the lowest (highest) masses at a given color or shape triaxiality.}
\label{fig:f5}
\end{figure*}

From Fig.~\ref{fig:f3}, we found that, in general, the mean of the $|\cos(\theta)|$ decreases with the value of x-axis. The transition from alignment to misalignment happen at  $\sim10^{9.4}\msun$ in the stellar mass, $\sim0.62$  in g–r color, $\sim0.4$ in triaxiality. We conclude that: \textit{for low-mass, blue, oblate galaxies, their spin tend to align with the slowest collapse direction, e3. However, a perpendicular trend was found in high-mass, red, prolate galaxies}. These trends are weakly dependent on the used scales to measure galaxy spin.

The results in Fig.~\ref{fig:f3} show that galaxy spin--LSS correlations are functions of galaxy mass, color, and triaxiality. Usually, the galaxy color and triaxiality are closed to the stellar mass such that low-mass galaxies are more likely to be blue and oblate. Fig.~\ref{fig:f4} we show the distribution of the galaxy stellar mass, color, and triaxiality from the simulation. No strong correlation between them was found. At a given stellar mass, there is a wide range of galaxy color and vice versa. To investigate which is the dominant effect in determining the spis--LSS correlation, in Fig.~\ref{fig:f5} we show the alignment angle $|\cos(\theta)|$ as a function of color (left panel) and triaxiality (right panel). At each color/triaxiality bin, we divide galaxies into three equal bin (in number) with increasing stellar masses, and we show the signal for the highest/lowest mass bins. Overall, galaxies in the lowest mass bins (blue lines) were always found to have $|\cos(\theta)|>0.5$, indicating an alignment signal, while the highest mass galaxies (red lines) have a misalignment. The trend of this alignment depends weakly on the color and triaxiality. Fig.~\ref{fig:f5} shows that it is the galaxy stellar mass that determines the spin--LSS correlation.

\begin{figure}
\plotone{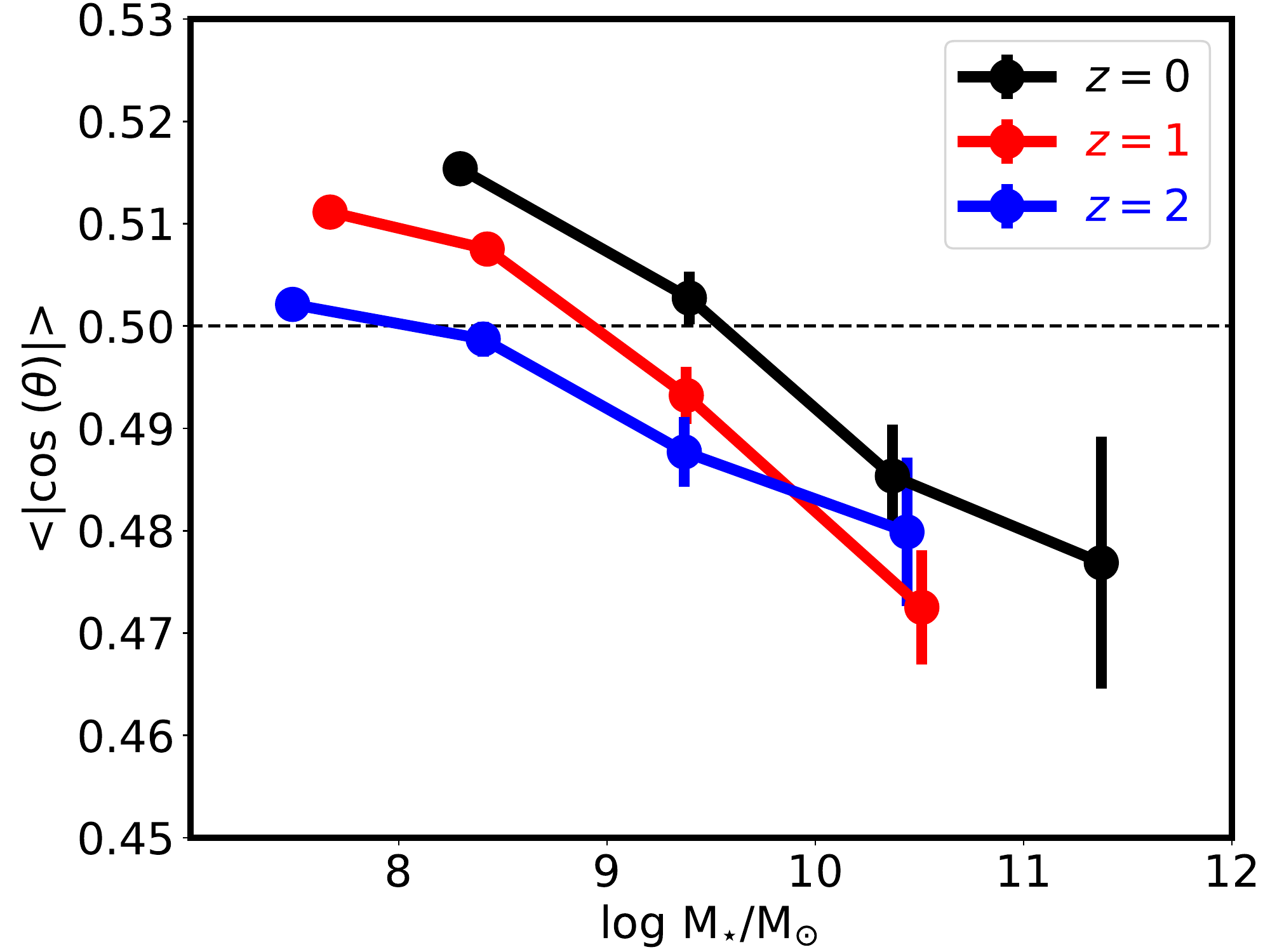}
\caption{Evolution of the spin--LSS correlations as a function of the galaxy stellar mass.}
\label{fig:f6}
\end{figure}

Fig.~\ref{fig:f6} shows the evolution of the spin--LSS alignment at z = 1, 2. As can be seen, the transition mass decreases with an increasing redshift. The stellar mass transition from alignment to misalignment happens around $10^{9.0}\msun$ at a redshift of $z=1$ and around $10^{8.0}\msun$ at a redshift of $z=2$. The evolution trend of the transition value of the stellar mass is consistent with the trend found with dark matter halos \citep[e.g.,][]{2018MNRAS.473.1562W}. However, it is in conflict with the value suggested by \cite{2014MNRAS.444.1453D} of $3\times10^{10}\msun$ at a redshift of $z=1.83$. We will discuss this in Section.~\ref{sec:con_dis}.

\section{Conclusion and Discussion}\label{sec:con_dis}
In this work, the galaxy spin--LSS correlation at $z = 0$ was investigated using the Illustris-1 hydrodynamic simulation. We have obtained these main results:
\begin{itemize}
\item Our work is the first to investigate the spin--LSS correlation at $z = 0$ using galaxies from cosmological hydrodynamical simulations. We found that, though weak but statistically significant, low-mass galaxies have their spins parallel to $\eee$ and massive galaxies have their spins perpendicular to $\eee$, where $\eee$ is the slowest collapse direction of the large-scale tidal field.
\item The spin--LSS correlation is correlated with the galaxy stellar mass, $g-r$ color and shape triaxiality. There is a large scatter between the galaxy mass and the color/triaxiality. We find that the stellar mass is the dominating factor in determining the spin--LSS correlation.
\item The transition position at redshift $z=0$ from parallel trend to perpendicular trend happen at $\sim10^{9.4}\msun$ in the stellar mass, $\sim0.62$ in g-r color, and $\sim0.4$ in triaxiality. The transition  stellar mass decreases with the increasing redshifts. 
\end{itemize}

Our results broadly consistent with previous results at $z=0$ using N-body simulations \citep[e.g.,][]{2007ApJ...655L...5A, 2007MNRAS.381...41H, 2013ApJ...762...72T, 2014MNRAS.443.1090F, 2017MNRAS.468L.123W, 2018MNRAS.473.1562W, 2018arXiv180500033G}. Our results are also the first work that uses simulated galaxies to verify the conclusion from the observational work verify the conclusion in the observations \citep{2013ApJ...775L..42T, 2009ApJ...706..747Z, 2016MNRAS.457..695P}. Additionally, we found that the galaxy spin--LSS alignment is primarily correlated with the stellar mass.

It is worth noting that observational measurement of a galaxy spin is tricky. The spin vectors of spiral galaxies are measured by the R.A., decl., and the inclination angle based on the projected axis ratio \citep{2006ApJ...640L.111T, 2007ApJ...671.1248L, 2012ApJ...744...82V}. For elliptical galaxies, the projected orientation of the minor axis is used with the assumption that it is aligned with their spin \citep{2013ApJ...775L..42T}. This is not accurate since there is usually a misalignment between the direction of the galaxy spin and the direction of the minor axis.

The origin of the spin--LSS correlation, especially the mass dependence, has prompted many studies in recent decades. Some of them \citep[e.g.,][]{2004MNRAS.352..376A, 2005ApJ...627..647B, 2005MNRAS.364..424W} have claimed that dark matter halo spin is a result of mergers and the conservation of angular momentum of accreted mass from larger scale. 

Some works \citep{2014MNRAS.443.1274L, 2014ApJ...786....8W, 2015ApJ...807...37S} have found that the mass accretion into halo is universal and that it is mainly along filaments. Such a scenario is able to explain the spin--LSS correlation for massive halo, but it fail to explain for low-mass halos. 
\cite{2015ApJ...813....6K} suggested that low-mass halo are preferentially fed by mass perpendicular to the $\eee$ direction and that massive halos are products of major mergers in filament though mass accretion along the filament direction. This non-universal mass accretion (two stage accretion pattern) can well explain the mass dependence of the spin--LSS correlation.

\cite{2017MNRAS.468L.123W} also have found that the spin--LSS correlation is closely related to halo formation time and the transition time when the halo environment changes. \cite{2018MNRAS.473.1562W} investigated the evolution of the halo spin--LSS correlation in details by claiming that, at early times, the spin of all halo progenitors is parallel with the LSS and that it evolves to parallel and perpendicular trends at $z=0$ depending on their mass growth history (Fig.4 in their paper). \cite{2014MNRAS.444.1453D} also show a similar picture in the Horizon-AGN simulation (Fig.8 in their paper), although their sample is limited to redshifts from $z=3.01$ to $z=1.23$. Therefore, the evolution of the galaxy spin--LSS after $z=1.23$ is still unknown.

The transition from the parallel trend to the perpendicular trend happens around $10^{8.0}\msun$ at redshift $z=2$ in this work. However, the transition mass at a redshift of $z = 1.83$, as claimed by \cite{2014MNRAS.444.1453D} is around $3\times10^{10}\msun$. The difference may come from the different method of LSS classification and the different physics governing star formation in the hydrodynamic simulation. The transition in the galaxy stellar mass at z = 0, 1, 2 from our work is not consistent with the empirical formula proposed by \cite{2012MNRAS.427.3320C} and \cite{2016IAUS..308..421P} (see the third paragraph in the Section 1). This indicates that the formation and evolution of the galaxy spin is different with that of the dark matter halo.

It is then interesting to ask how well the spin of a galaxy is correlated with its host dark matter halo and its large-scale cosmic-web environment. The tentative consistency comes with the assumption that the spin of a galaxy is similar to that of its host halo, which does not necessary hold true. The size of a galaxy is typically 10\% of the virial radius of its host halo. It is not obvious that the spin of the galaxy, which occupies only the very central part of the halo, should necessarily follow the spin of the entire host halo. For example, \cite{2016MNRAS.460.3772S} found that the center galaxy is better aligned with the inner $\rm 10 \ kpc$ of the host halo. \cite{2017MNRAS.472.1163C}  claimed that the mean misalignment angle between the minor axis of a galaxy and its halo is a strong function of the halo mass, but when the galaxies are divided into disks and ellipticals, there is a significant residual in this relationship. The question then can be rephrased as the follows: should the spin of the inner part of a halo be the same as that of the halo as a whole? This call for more studies, which we will investigate in future work.

\acknowledgments
{\bf Acknowledgement:}
We thank the anonymous referee for careful reading and constructive suggestions that improved the presentation of our paper. We thanks the Illustris team make the simulation data publicly available. The work  is supported  by the NSFC (No.  11333008), the  973 program  (No.  2015CB857003), the NSFC (No.11703091), the NSF of Jiangsu Province (No. BK20140050). QG acknowledges supports by the NSFC (No. 11743003). 
\bibliographystyle{apj}

\begin{thebibliography}{}

\bibitem[\protect\citeauthoryear{Arag{\'o}n-Calvo et al.}{2007a}]{2007A&A...474..315A} Arag{\'o}n-Calvo M.~A., Jones B.~J.~T., van de Weygaert R., van der Hulst J.~M., 2007a, A\&A, 474, 315 
\bibitem[\protect\citeauthoryear{Arag{\'o}n-Calvo et al.}{2007b}]{2007ApJ...655L...5A} Arag{\'o}n-Calvo M.~A., van de Weygaert R., Jones B.~J.~T., van der Hulst J.~M., 2007b, ApJ, 655, L5 
\bibitem[\protect\citeauthoryear{Aragon-Calvo \& Yang}{2014}]{2014MNRAS.440L..46A} Aragon-Calvo M.~A., Yang L.~F., 2014, MNRAS, 440, L46 
\bibitem[\protect\citeauthoryear{Aubert, Pichon, \& Colombi}{2004}]{2004MNRAS.352..376A} Aubert D., Pichon C., Colombi S., 2004, MNRAS, 352, 376 
\bibitem[\protect\citeauthoryear{Bailin \& Steinmetz}{2005}]{2005ApJ...627..647B} Bailin J., Steinmetz M., 2005, ApJ, 627, 647 
\bibitem[\protect\citeauthoryear{Barnes \& Efstathiou}{1987}]{1987ApJ...319..575B} Barnes J., Efstathiou G., 1987, ApJ, 319, 575
\bibitem[\protect\citeauthoryear{Chisari et al.}{2017}]{2017MNRAS.472.1163C} Chisari N.~E., et al., 2017, MNRAS, 472, 1163
\bibitem[\protect\citeauthoryear{Codis et al.}{2012}]{2012MNRAS.427.3320C} Codis S., Pichon C., Devriendt J., Slyz A., Pogosyan D., Dubois Y., Sousbie T., 2012, MNRAS, 427, 3320 
\bibitem[\protect\citeauthoryear{Davis et al.}{1985}]{1985ApJ...292..371D} Davis M., Efstathiou G., Frenk C.~S., White S.~D.~M., 1985, ApJ, 292, 371 
\bibitem[\protect\citeauthoryear{Dolag et al.}{2009}]{2009MNRAS.399..497D} Dolag K., Borgani S., Murante G., Springel V., 2009, MNRAS, 399, 497 
\bibitem[\protect\citeauthoryear{Dubois et al.}{2014}]{2014MNRAS.444.1453D} Dubois Y., et al., 2014, MNRAS, 444, 1453 
\bibitem[\protect\citeauthoryear{Forero-Romero, Contreras, \& Padilla}{2014}]{2014MNRAS.443.1090F} Forero-Romero J.~E., Contreras S., Padilla N., 2014, MNRAS, 443, 1090 
\bibitem[\protect\citeauthoryear{Ganeshaiah Veena et al.}{2018}]{2018arXiv180500033G} Ganeshaiah Veena P., Cautun M., van de Weygaert R., Tempel E., Jones B.~J.~T., Rieder S., Frenk C.~S., 2018, arXiv, arXiv:1805.00033 
\bibitem[\protect\citeauthoryear{Hahn et al.}{2007}]{2007MNRAS.381...41H} Hahn O., Carollo C.~M., Porciani C., Dekel A., 2007, MNRAS, 381, 41 
\bibitem[\protect\citeauthoryear{Hinshaw et al.}{2013}]{2013ApJS..208...19H} Hinshaw G., et al., 2013, ApJS, 208, 19 
\bibitem[\protect\citeauthoryear{Hoffman et al.}{2012}]{2012MNRAS.425.2049H} Hoffman Y., Metuki O., Yepes G., Gottl{\"o}ber S., Forero-Romero J.~E., Libeskind N.~I., Knebe A., 2012, MNRAS, 425, 2049 
\bibitem[\protect\citeauthoryear{Hoyle}{1951}]{1951pca..conf..195H} Hoyle F., 1951, pca..conf, 195 
\bibitem[\protect\citeauthoryear{Huchra et al.}{2012}]{2012ApJS..199...26H} Huchra J.~P., et al., 2012, ApJS, 199, 26 
\bibitem[\protect\citeauthoryear{Kang \& Wang}{2015}]{2015ApJ...813....6K} Kang X., Wang P., 2015, ApJ, 813, 6 
\bibitem[\protect\citeauthoryear{Lee \& Erdogdu}{2007}]{2007ApJ...671.1248L} Lee J., Erdogdu P., 2007, ApJ, 671, 1248 
\bibitem[\protect\citeauthoryear{Libeskind et al.}{2013}]{2013MNRAS.428.2489L} Libeskind N.~I., Hoffman Y., Forero-Romero J., Gottl{\"o}ber S., Knebe A., Steinmetz M., Klypin A., 2013, MNRAS, 428, 2489 
\bibitem[\protect\citeauthoryear{Libeskind et al.}{2014}]{2014MNRAS.443.1274L} Libeskind N.~I., Knebe A., Hoffman Y., Gottl{\"o}ber S., 2014, MNRAS, 443, 1274 
\bibitem[\protect\citeauthoryear{Libeskind et al.}{2018}]{2018MNRAS.473.1195L} Libeskind N.~I., et al., 2018, MNRAS, 473, 1195 
\bibitem[\protect\citeauthoryear{Pahwa et al.}{2016}]{2016MNRAS.457..695P} Pahwa I., et al., 2016, MNRAS, 457, 695 
\bibitem[\protect\citeauthoryear{Peebles}{1969}]{1969ApJ...155..393P} Peebles P.~J.~E., 1969, ApJ, 155, 393 
\bibitem[\protect\citeauthoryear{Shao et al.}{2016}]{2016MNRAS.460.3772S} Shao S., Cautun M., Frenk C.~S., Gao L., Crain R.~A., Schaller M., Schaye J., Theuns T., 2016, MNRAS, 460, 3772
\bibitem[\protect\citeauthoryear{Pichon et al.}{2016}]{2016IAUS..308..421P} Pichon C., Codis S., Pogosyan D., Dubois Y., Desjacques V., Devriendt J., 2016, IAUS, 308, 421 
\bibitem[\protect\citeauthoryear{Shi, Wang, \& Mo}{2015}]{2015ApJ...807...37S} Shi J., Wang H., Mo H.~J., 2015, ApJ, 807, 37 
\bibitem[\protect\citeauthoryear{Springel}{2010}]{2010MNRAS.401..791S} Springel V., 2010, MNRAS, 401, 791 
\bibitem[\protect\citeauthoryear{Springel et al.}{2001}]{2001MNRAS.328..726S} Springel V., White S.~D.~M., Tormen G., Kauffmann G., 2001, MNRAS, 328, 726 
\bibitem[\protect\citeauthoryear{Tempel et al.}{2014}]{2014MNRAS.438.3465T} Tempel E., Stoica R.~S., Mart{\'{\i}}nez V.~J., Liivam{\"a}gi L.~J., Castellan G., Saar E., 2014, MNRAS, 438, 3465 
\bibitem[\protect\citeauthoryear{Tempel, Stoica, \& Saar}{2013}]{2013MNRAS.428.1827T} Tempel E., Stoica R.~S., Saar E., 2013, MNRAS, 428, 1827 
\bibitem[\protect\citeauthoryear{Tempel \& Libeskind}{2013}]{2013ApJ...775L..42T} Tempel E., Libeskind N.~I., 2013, ApJ, 775, L42 
\bibitem[\protect\citeauthoryear{Trowland, Lewis, \& Bland-Hawthorn}{2013}]{2013ApJ...762...72T} Trowland H.~E., Lewis G.~F., Bland-Hawthorn J., 2013, ApJ, 762, 72 
\bibitem[\protect\citeauthoryear{Trujillo, Carretero, \& Patiri}{2006}]{2006ApJ...640L.111T} Trujillo I., Carretero C., Patiri S.~G., 2006, ApJ, 640, L111 
\bibitem[\protect\citeauthoryear{Varela et al.}{2012}]{2012ApJ...744...82V} Varela J., Betancort-Rijo J., Trujillo I., Ricciardelli E., 2012, ApJ, 744, 82 
\bibitem[\protect\citeauthoryear{Vogelsberger et al.}{2014}]{2014MNRAS.444.1518V} Vogelsberger M., et al., 2014, MNRAS, 444, 1518 
\bibitem[\protect\citeauthoryear{Wang et al.}{2005}]{2005MNRAS.364..424W} Wang H.~Y., Jing Y.~P., Mao S., Kang X., 2005, MNRAS, 364, 424 
\bibitem[\protect\citeauthoryear{Wang \& Kang}{2018}]{2018MNRAS.473.1562W} Wang P., Kang X., 2018, MNRAS, 473, 1562 
\bibitem[\protect\citeauthoryear{Wang \& Kang}{2017}]{2017MNRAS.468L.123W} Wang P., Kang X., 2017, MNRAS, 468, L123 
\bibitem[\protect\citeauthoryear{Wang et al.}{2014}]{2014ApJ...786....8W} Wang Y.~O., Lin W.~P., Kang X., Dutton A., Yu Y., Macci{\`o} A.~V., 2014, ApJ, 786, 8 
\bibitem[\protect\citeauthoryear{White}{1984}]{1984ApJ...286...38W} White S.~D.~M., 1984, ApJ, 286, 38
\bibitem[\protect\citeauthoryear{Zel'dovich}{1970}]{1970A&A.....5...84Z} Zel'dovich Y.~B., 1970, A\&A, 5, 84  
\bibitem[\protect\citeauthoryear{Zhang et al.}{2009}]{2009ApJ...706..747Z} Zhang Y., Yang X., Faltenbacher A., Springel V., Lin W., Wang H., 2009, ApJ, 706, 747 
\bibitem[\protect\citeauthoryear{Zhang et al.}{2015}]{2015ApJ...798...17Z} Zhang Y., Yang X., Wang H., Wang L., Luo W., Mo H.~J., van den Bosch F.~C., 2015, ApJ, 798, 17 

\end{thebibliography}

\end{document}